# Estimation of Domain Size in Nano Ferroelectrics from NMR $T_1$ Measurements

*Alok Shukla*[1], *Mrignayani Kotecha*[2*] *and Lakshman Pandey*[3]

1 Department of Physics, Rani Durgavati University, Jabalpur-482001, (M.P.), India.

2 Department of Chemistry, University of Illinois at Chicago, Chicago, 60607, USA

## Abstract

The spin lattice relaxation of I=3/2 quadrupolar spin system due to domain walls in order-disorder ferroelectrics has been studied and a general method is proposed for the measurement of domain width in nano ferroelectrics. Based on the fact that electric polarization undergoes spiral orientation as one moves from one domain to the other, it is assumed that at low temperatures the spins at and near domain walls undergo relaxation due to possible easy reorientation of electric polarization in domain walls even though such a relaxation in the main body of the domain has almost ceased. The spins present inside the domain undergo relaxation through transfer of magnetization to the domain walls through a spin diffusion process by nearest neighbour interaction. Rate equations for spin populations are formed by representing the ferroelectric domain by a one-dimensional chain of equidistant spins having dipolar coupling. Spin populations are calculated as a function of time for different ratios of quadrupolar to dipolar transition probabilities for a sample subjected to selective rf pulse. Expression for spin- lattice relaxation time $T_1$ is derived in terms of domain width and ratio of quadrupolar to dipolar transition probabilities. It is found that the domain width can be estimated provided the value of spin lattice relaxation time $T_1$ is known for the corresponding crystal with normal sized grains. The results are quite general and can be applied to any order disorder ferroelectric with nano sized domains and having spin I=3/2 nuclei.

PACS: 61.46.-w, 76.60.-k, 77.80.-e, 77.84.-s, 77.80.Dj

## 1. Introduction:

Ferroelectrics are a very important class of materials having wide variety of applications in various technological devices such as electro–optic materials, infrared sensors, ultrasonic systems, actuators, electric field and strain sensors, nonvolatile memory devices etc. [1-7]. A material is said to be ferroelectric when it has two or more orientational states of electric polarization and can be reoriented from one state to another by an electric field. If the spontaneous polarization arises due to the ordering of ions or some group of ions, then the ferroelectric is said to be order–disorder ferroelectric [1-3]. The thermal motions tend to destroy the ferroelectric order and ferroelectricity usually disappears beyond a certain temperature $T_c$, called the transition temperature. Below $T_c$, a ferroelectric material comprises regions of uniform polarization, called domains. Within each domain, the polarization is in the same direction, but in the adjacent domain it is in different direction. The region joining two adjacent domains is called domain–wall. If the spontaneous polarizations in the adjacent domains are in opposite directions, the domains are called $180^o$ domains and the region joining two $180^o$ domains are called $180^o$domain-walls. Figure 1 shows the schematic diagram of 180° domain-domain wall structure. The technological applications of a ferroelectric greatly depend upon its domain structure and behavior and shape of hysteresis loop that in turn is governed by how fast the domains can be switched from one direction to the other. The switching process involves building up of the favorable domains at the expense of the unfavorable ones starting from nucleation and growth at the domain-walls [2]. Also the properties of a ferroelectric tend to change over period of time due to gradual build up of inhibiting structure at domain walls reducing their mobility [1, 2].

Due to miniaturization trend in device size, the size dependent evolution of ferroelectricity in nano crystalline and thin films samples has been focus of many recent research efforts [5, 8-20]. There have been reports indicating that some ferroelectrics loose their ferroelectricity when the grain size is reduced to nanometer range. However, several investigations on nano-sized ferroelectric ceramics have been carried out [21-24]. The details of static and dynamics of nano-ferroelectrics have been recently summarized by Scott [25]. Single crystalline ferroelectric nano wires and nano tubes have been produced with retention time for the induced polarization exceeding several days [15]. Nano sized nonvolatile polarization domains can be induced on these nano-wires suggesting that ferroelectric nano wires may be used to fabricate non volatile memory devices with an integration density approaching 1 terrabit/cm$^2$. The potential for application as non volatile random access memory has stimulated great interest in the integration of ferroelectric thin films and nano structures [18-20, 26]. For ultra high density integration of ferroelectric memories the investigation of size effect and the estimation of domain width become extremely important [26].

The ferroelectric properties of thin films samples are studied by using ultra high vacuum Scanned Probe Microscope (SPM). The written polarization is read by using electrostatic force microscopy (EFM) by measuring the shift in the resonance frequency of a SPM cantilever while scanning it with a small tip voltage [15]. The shift is directly proportional to the electrostatic force felt by the tip and thus to the magnitude of the electric polarization of the nano wire. A plot of the shift as a function of tip position provides a spatial map of electric polarization direction on the nano wire. It has been shown that measured domain size as observed through the EFM is limited by the tip-sample distance (~few tens of nm) due to long ranged nature of electrostatic interaction. Similarly it has been demonstrated that induced ferroelectric domains as applied through the AFM tip formed information bits with size of 60 nm diameter in PZT. The formed bits were recorded back with high spatial resolution of ~ 10 nm. Also the polarization retention time is dependent on size of domains [27]. Thus the determination of the size of domains is very crucial for a strategy to achieve maximum integration density.

Study of ferroelectric domains and domain walls including the local dynamics has been drawing considerable attention of research workers in the past as well as in recent years [4, 5, 28-36]. For this various techniques [1] such as optical birefringence, second harmonic generation, electron microscopy, chemical etching, X-ray topography, U.V. photoemission, electrostatic force microscopy, atomic force microscopy etc. have been used for different materials. Nuclear magnetic resonance (NMR) has been a very powerful tool for studying the local environment [1, 37-39].

Nuclei with spin $I \geq 1$ possess electric quadrupole moment in addition to magnetic dipole moment. The dipole moments interact with the local magnetic fields whereas the electric quadrupole moments interact with the electric field gradients. As a result these nuclei can sense subtle changes taking place in the local environment and these changes are recorded through the NMR line widths and relaxation times. The NMR studies of nuclei with $I \geq 1$ have therefore proved to be a very powerful tool for the study of local structure and dynamics including phase transitions. A large number of reports are available on successful use of NMR for such investigations in ferroelectrics; a few illustrative ones are given in the reference [40-46]. However, the possible effect of domain walls on the spin relaxation does not seem to have been studied so far in literature for any ferroelectric system except some preliminary work by Kotecha and Pandey [47]. In this paper we present a theoretical study of the nuclear spin relaxation of spin I=3/2 quadrupolar system due to domain walls in order disorder ferroelectrics and its possible application for the determination of domain size in nano ferroelectrics. The necessary rate equations for the change of population of various levels in I=3/2 spin system is developed and solved in the next section followed by results and discussion.

## 2. Model for domain size calculation in nano-ferroelectrics

Let us consider a ferroelectric material of order disorder type possessing I=3/2 nuclei. We assume that a $180^0$ domain can be represented by a one-dimensional array of equidistant nuclei situated at …x-2$a$, x-$a$, x, x+$a$, x+2$a$ ... as shown in figure 2. In an external magnetic field, the I=3/2 spins would have four Zeeman levels [37, 38] with populations $n_{3/2}$, $n_{1/2}$, $n_{-1/2}$ and $n_{-3/2}$ corresponding to the quantum numbers m=3/2, 1/2, -1/2, -3/2. We assume that each atom interacts with its nearest neighbours only and the quadrupole-coupling constant is such that it gives rise to well resolved NMR spectrum corresponding to the centre line (1/2↔-1/2) and satellite transitions (±3/2↔±1/2).

The rate of change of deviations of populations from thermal equilibrium values can be written as [37, 38, 47-49]

$$\begin{aligned}\frac{\partial n_{3/2}(x,t)}{\partial t} = & -2n_{3/2}(x,t)W_1^1 + 2n_{1/2}(x,t)\,W_1^1 - \frac{1}{N}\,n_{3/2}(x,t)\,n_{1/2}(x+a,t)\,W_0^1 \\ & + \frac{1}{N}\,n_{1/2}(x,t)\,n_{3/2}(x+a,t)\,W_0^1 - \frac{1}{N}\,n_{3/2}(x,t)\,n_{1/2}(x-a,t)\,W_0^1 \\ & + \frac{1}{N}\,n_{1/2}(x,t)\,n_{3/2}(x-a,t)\,W_0^1 - \frac{1}{N}\,n_{3/2}(x,t)\,n_{3/2}(x+a,t)\,W_2^1 \\ & + \frac{1}{N}\,n_{1/2}(x,t)\,n_{1/2}(x+a,t)\,W_2^1 - \frac{1}{N}\,n_{3/2}(x,t)\,n_{3/2}(x-a,t)\,W_2^1 \\ & + \frac{1}{N}\,n_{1/2}(x,t)\,n_{1/2}(x-a,t)\,W_2^1\end{aligned}$$

$$\begin{aligned}\frac{\partial n_{1/2}(x,t)}{\partial t} = & \; 2n_{3/2}(x,t)W_1^1 - 2n_{1/2}(x,t)\,W_1^1 - 2n_{1/2}(x,t)W_1^0 + 2n_{-1/2}(x,t)\,W_1^0 \\ & - \frac{1}{N}\,n_{3/2}(x-a,t)\,n_{1/2}(x,t)\,W_0^1 - \frac{1}{N}\,n_{1/2}(x,t)\,n_{3/2}(x+a,t)\,W_0^1 \\ & + \frac{1}{N}\,n_{3/2}(x,t)\,n_{1/2}(x-a,t)\,W_0^1 + \frac{1}{N}\,n_{3/2}(x,t)\,n_{1/2}(x+a,t)\,W_0^1 \\ & + \frac{1}{N}\,n_{-1/2}(x,t)\,n_{1/2}(x+a,t)\,W_0^0 - \frac{1}{N}\,n_{1/2}(x,t)\,n_{-1/2}(x-a,t)\,W_0^0 \\ & + \frac{1}{N}\,n_{-1/2}(x,t)\,n_{1/2}(x-a,t)\,W_0^0 - \frac{1}{N}\,n_{1/2}(x,t)\,n_{-1/2}(x+a,t)\,W_0^0 \\ & + \frac{1}{N}\,n_{3/2}(x,t)\,n_{3/2}(x+a,t)\,W_2^1 - \frac{1}{N}\,n_{1/2}(x,t)\,n_{1/2}(x+a,t)\,W_2^1 \\ & + \frac{1}{N}\,n_{3/2}(x,t)\,n_{3/2}(x-a,t)\,W_2^1 - \frac{1}{N}\,n_{1/2}(x,t)\,n_{1/2}(x-a,t)\,W_2^1 \\ & - \frac{1}{N}\,n_{1/2}(x,t)\,n_{1/2}(x+a,t)\,W_2^0 + \frac{1}{N}\,n_{-1/2}(x,t)\,n_{-1/2}(x+a,t)\,W_2^0 \\ & - \frac{1}{N}\,n_{1/2}(x,t)\,n_{1/2}(x-a,t)\,W_2^0 + \frac{1}{N}\,n_{-1/2}(x,t)\,n_{-1/2}(x-a,t)\,W_2^0\end{aligned}$$

$$
\begin{aligned}
\frac{\partial n_{-1/2}(x,t)}{\partial t} = \; & 2n_{1/2}(x,t)W_1^0 - 2n_{-1/2}(x,t)\, W_1^0 \; - \; 2n_{-1/2}(x,t)W_1^{-1} + 2n_{-3/2}(x,t)\, W_1^{-1} \\
& - \frac{1}{N} n_{1/2}(x-a,t)\, n_{-1/2}(x,t)\, W_0^0 \quad - \frac{1}{N}\, n_{-1/2}(x,t)\, n_{1/2}(x+a,t)\, W_0^0 \\
& + \frac{1}{N}\, n_{1/2}(x,t)\, n_{-1/2}(x-a,t)\, W_0^0 \quad + \frac{1}{N}\, n_{1/2}(x,t)\, n_{-1/2}(x+a,t)\, W_0^0 \\
& + \frac{1}{N}\, n_{-3/2}(x,t)\, n_{-1/2}(x+a,t)\, W_0^{-1} - \frac{1}{N}\, n_{-1/2}(x,t)\, n_{-3/2}(x-a,t)\, W_0^{-1} \\
& + \frac{1}{N}\, n_{-3/2}(x,t)\, n_{-1/2}(x-a,t)\, W_0^{-1} - \frac{1}{N}\, n_{-1/2}(x,t)\, n_{-3/2}(x+a,t)\, W_0^{-1} \\
& + \frac{1}{N}\, n_{1/2}(x,t) n_{1/2}(x+a,t)\, W_2^0 \quad - \frac{1}{N}\, n_{-1/2}(x,t)\, n_{-1/2}(x+a,t)\, W_2^0 \\
& + \frac{1}{N}\, n_{1/2}(x,t)\, n_{1/2}(x-a,t)\, W_2^0 \quad - \frac{1}{N}\, n_{-1/2}(x,t)\, n_{-1/2}(x-a,t)\, W_2^0 \\
& + \frac{1}{N}\, n_{1/2}(x,t)\, n_{1/2}(x-a,t)\, W_2^0 \quad - \frac{1}{N}\, n_{-1/2}(x,t)\, n_{-1/2}(x-a,t)\, W_2^0 \\
& - \frac{1}{N}\, n_{-1/2}(x,t)\, n_{-1/2}(x+a,t)\, W_2^{-1} + \frac{1}{N}\, n_{-3/2}(x,t)\, n_{-3/2}(x+a,t)\, W_2^{-1} \\
& - \frac{1}{N}\, n_{-1/2}(x,t)\, n_{-1/2}(x-a,t)\, W_2^{-1} + \frac{1}{N}\, n_{-3/2}(x,t)\, n_{-3/2}(x-a,t)\, W_2^{-1}
\end{aligned}
$$

$$
\begin{aligned}
\frac{\partial n_{-3/2}(x,t)}{\partial t} = \; & 2n_{-1/2}(x,t)\, W_1^{-1} \; - \; 2n_{-3/2}(x,t)\, W_1^{-1} \\
& + \frac{1}{N} n_{-1/2}(x,t) n_{-3/2}(x+a,t)\, W_0^{-1} - \frac{1}{N} n_{-3/2}(x,t) n_{-1/2}(x+a,t)\, W_0^{-1} \\
& + \frac{1}{N} n_{-1/2}(x,t) n_{-3/2}(x-a,t)\, W_0^{-1} - \frac{1}{N} n_{-3/2}(x,t) n_{-1/2}(x-a,t)\, W_0^{-1} \\
& + \frac{1}{N} n_{-1/2}(x,t) n_{-1/2}(x+a,t)\, W_2^{-1} - \frac{1}{N} n_{-3/2}(x,t) n_{-3/2}(x+a,t)\, W_2^{-1} \\
& + \frac{1}{N} n_{-1/2}(x,t) n_{-1/2}(x-a,t)\, W_2^{-1} - \frac{1}{N} n_{-3/2}(x,t) n_{-3/2}(x-a,t)\, W_2^{-1}
\end{aligned}
$$

(1)

where $N = n_{3/2} + n_{1/2} + n_{-1/2} + n_{-3/2}$

and $W^1{}_0$, $W^0{}_0$, $W^{-1}{}_0$ are the transition probabilities of 3/2 $\leftrightarrow$ 1/2, 1/2 $\leftrightarrow$ -1/2, -1/2 $\leftrightarrow$ -3/2 levels respectively for the case where one spin is undergoing an upward transitions while the other spin undergoes downward transition (usually called flip-flop term) [37,38]. The probabilities $W^1{}_2$, $W^0{}_2$, $W^{-1}{}_2$ represent simultaneous upward (or down ward) flip of the pair of spins. Similarly $W^1{}_1$, $W^0{}_1$, $W^{-1}{}_1$ represent the single spin transition probability for the spin-pairs and counted twice for each pair of spins. Here we have used the same symbols and notations as used earlier in Ref. [47].

Defining

$N_{+1} = n_{3/2} - n_{1/2}$

$N_0 = n_{+1/2} - n_{-1/2}$

$N_{-1} = n_{-1/2} - n_{-3/2}$

and assuming that

$W_0^i \equiv W_0^i(x, x+a) \equiv W_0^i(x, x-a)$ (for $i$ = 1, 0, -1)

and $W_2^i \equiv W_2^i(x, x+a) \equiv W_2^i(x, x-a)$

We can rewrite eq. (1) as

$$\frac{\partial N_{+1}}{\partial t}(x,t) = -2\rho N_{+1}(x,t) - 2\sigma N_{+1}(x+a,t) - 2\sigma N_{+1}(x-a,t) + \rho' N_0(x,t) + \sigma' N_0(x+a,t) + \sigma' N_0(x-a,t)$$

$$\frac{\partial N_0}{\partial t}(x,t) = \rho N_{+1}(x,t) + \sigma N_{+1}(x+a,t) + \sigma N_{+1}(x-a,t) - 2\rho' N_0(x,t) - 2\sigma' N_0(x+a,t) - 2\sigma' N_0(x-a,t) + \rho'' N_{-1}(x,t) + \sigma'' N_{-1}(x+a,t) + \sigma'' N_{-1}(x-a,t)$$

$$\frac{\partial N_{-1}}{\partial t}(x,t) = \rho' N_0(x,t) + \sigma' N_0(x+a,t) + \sigma' N_0(x-a,t) - 2\rho'' N_{-1}(x,t) - 2\sigma'' N_{-1}(x+a,t) - 2\sigma'' N_{-1}(x-a,t)$$

where

$$\rho = 2W_1^1 + \frac{W_0^1}{2} + \frac{W_2^1}{2}, \qquad \sigma = \frac{W_2^1 - W_0^1}{4}$$

$$\rho' = 2W_1^0 + \frac{W_0^0}{2} + \frac{W_2^0}{2}, \qquad \sigma' = \frac{W_2^0 - W_0^0}{4}$$

$$\rho'' = 2W_1^{-1} + \frac{W_0^{-1}}{2} + \frac{W_2^{-1}}{2}, \qquad \sigma'' = \frac{W_2^{-1} - W_0^{-1}}{4} \tag{2}$$

Expanding the terms $N_{+1}(x+a, t)$, $N_{-1}(x+a, t)$, $N_0(x+a, t)$, $N_0(x-a, t)$, $N_{-1}(x-a, t)$, $N_{+1}(x-a, t)$ using Taylor series expansion, and retaining the terms up to the second derivative, and taking

$\rho + 2\sigma = a_1$, $\rho' + 2\sigma' = a_2$, $\rho'' + 2\sigma'' = a_3$,

$D_1 = -\sigma a^2$, $D_2 = -\sigma' a^2$, $D_3 = -\sigma'' a^2$,

the set of eq. (2) reduces to

$$\frac{\partial N_{+1}}{\partial t}(x,t) = -2a_1 N_{+1}(x,t) + 2D_1 \frac{\partial^2}{\partial x^2} N_{+1}(x,t) + a_2 N_0(x,t) - D_2 \frac{\partial^2}{\partial x^2} N_0(x,t)$$

$$\frac{\partial N_0}{\partial t}(x,t) = a_1 N_{+1}(x,t) - D_1 \frac{\partial^2}{\partial x^2} N_{+1}(x,t) - 2a_2 N_0(x,t) + 2D_2 \frac{\partial^2}{\partial x^2} N_0(x,t) + a_3 N_{-1}(x,t) - D_3 \frac{\partial^2}{\partial x^2} N_{-1}(x,t)$$

$$\frac{\partial N_{-1}}{\partial t}(x,t) = a_2 N_0(x,t) - D_2 \frac{\partial^2}{\partial x^2} N_0(x,t) - 2a_3 N_{-1}(x,t) + 2D_3 \frac{\partial^2}{\partial x^2} N_{-1}(x,t)$$

(3)

These coupled simultaneous differential equations represent spin diffusion and can be solved by using Laplace Transform [50, 51]. Taking the Laplace Transform of eq. (3) over the variable $t$ and writing

$Z_{+1} = \mathcal{L}\{ N_{+1}(x, t)\}$, $Z_0 = \mathcal{L}\{ N_0(x, t)\}$, $Z_{-1} = \mathcal{L}\{ N_{-1}(x, t)\}$,

we get

$$4C_2 \frac{\partial^2 Z_{+1}}{\partial x^2} + (4C_1 + 3s) Z_{+1} + 2sZ_0 + sZ_{-1} = k_1$$

$$2C_4 \frac{\partial^2 Z_0}{\partial x^2} + s\,Z_{+1} + \left(2s + 2C_3\right) Z_0 + sZ_{-1} = k_2$$

$$4C_6 \frac{\partial^2 Z_{-1}}{\partial x^2} + s\,Z_{+1} + 2s\,Z_0 + \left(4C_5 + 3s\right) Z_{-1} = k_3 \qquad (4)$$

with

$k_1 = 3\,N_{+1}(x, 0) + N_{-1}(x, 0) + 2N_0(x, 0)$

$k_2 = N_{+1}(x, 0) + N_{-1}(x, 0) + 2\,N_0(x, 0)$

$k_3 = N_{+1}(x, 0) + 3N_{-1}(x, 0) + 2N_0(x, 0)$

and $C_1 = 2W_1^1 + W_2^1$ , $\qquad C_2 = -\frac{a^2}{4}\left(W_0^1 - W_2^1\right)$

$C_3 = 2W_1^0 + W_2^0$ , $\qquad C_4 = -\frac{a^2}{4}\left(W_0^0 - W_2^0\right)$

$C_5 = 2W_1^{-1} + W_2^{-1}$, $\qquad C_6 = -\frac{a^2}{4}\left(W_0^{-1} - W_2^{-1}\right)$

For the given I = 3/2 system, the value of the probabilities can be written as [38]

$$W_0^1 = \frac{9}{8}\frac{A_0^2}{\hbar^2}\left(1 - 3\cos^2\theta\right)^2 \tau_c \qquad W_1^1 = \frac{27}{8}\frac{A_0^2}{\hbar^2}\sin^2\theta\cos^2\theta\,\tau_c$$

$$W_0^0 = 2\frac{A_0^2}{\hbar^2}\left(1 - 3\cos^2\theta\right)^2 \tau_c \qquad W_1^0 = \frac{9}{2}\frac{A_0^2}{\hbar^2}\sin^2\theta\cos^2\theta\,\tau_c$$

$$W_0^{-1} = \frac{9}{8}\frac{A_0^2}{\hbar^2}\left(1 - 3\cos^2\theta\right)^2 \tau_c \qquad W_1^{-1} = \frac{243}{8}\frac{A_0^2}{\hbar^2}\sin^2\theta\cos^2\theta\,\tau_c$$

$$W_2^1 = \frac{81}{8}\frac{A_0^2}{\hbar^2}\sin^4\theta\,\tau_c$$

$$W_2^0 = 18\frac{A_0^2}{\hbar^2}\sin^4\theta\,\tau_c$$

$$W_2^{-1} = \frac{81}{8}\frac{A_0^2}{\hbar^2}\sin^4\theta\,\tau_c$$

where $A_0 = \gamma_I\gamma_I\hbar^2 / r^3$, θ is the polar angle of the radius vector joining two nuclei with respect to the external magnetic field and $\tau_c$ is the correlation time.

The eqs. (4) were difficult to solve for the general case. So we made a simplifying assumption that the direction of the external static magnetic field used in a NMR experiment is at right angles to the array of spins, so that θ is equal to $90^0$. We further assume that there is no cross-relaxation between the satellite and centre line transitions and only flip-flop terms are retained. We therefore set $C_1$, $C_3$ and $C_5$ in eq. (4) equal to zero. The eq. (4) can then be written as

$$Z''_{+1} = -\frac{1}{9|b|}\left[k_1 - 3s\, Z_{+1} - 2sZ_0 - sZ_{-1}\right]$$

$$Z''_{0} = -\frac{1}{8|b|}\left[k_2 - s\, Z_{+1} - 2sZ_0 - sZ_{-1}\right]$$

$$Z''_{-1} = -\frac{1}{9|b|}\left[k_3 - s\, Z_{+1} - 2sZ_0 - 3sZ_{-1}\right] \qquad (5)$$

where $|b| = \frac{a^2 W_{00}}{16}$, $W_{00} = W_0^0 = \left(\frac{2A_0^2}{\hbar^2}\right)\left(1 - 3\cos^2\theta\right)$

'*s*' being the Laplace variable (see Appendix-I)

The set of eq. (5) can be solved for different initial and boundary conditions. For an easy comparison of the results with the generally performed pulsed NMR relaxation measurements and also to study the domain-wall effects, we consider the following situations.

## 3. Boundary conditions for NMR relaxation experiments and effect of domain wall at low temperatures

The centre line of the quadrupole split well resolved spectrum of I = 3/2 system is subjected to a selective radio frequency NMR pulse for a duration such that a fraction α of spins flip from the lower state I = 1/2 to the higher state I = -1/2 and the population differences become

$$N_0(x, 0) = -2\alpha, \quad N_{-1}(x, 0) = \alpha, \quad N_{+1}(x, 0) = \alpha \qquad (6)$$

The time t = 0 corresponds to the end of the pulse. In forthcoming calculation a π/2 pulse would be considered so that the value of α would be 0.5. A $180^o$ domain with the domain-wall at its end is considered. The origin of the coordinates, x = 0, is taken at the domain-wall and the wall is taken to be thin.

It is known that spin lattice relaxation of quadrupolar nuclei in ferroelectrics usually occurs predominantly through interaction of quadrupole moment with the fluctuation electric field gradient that are created by flipping motions of group of ions (e.g. $NO_2$ in $NaNO_2$). Also it was shown by Hughes and Pandey [44, 45] that in order-disorder ferroelectrics the electric polarization undergoes a spiral orientation as one move from one domain to the other. If we visualize the whole sample to be made up of thin slices, then it means that the polarization in adjacent slices in the larger body of the domains are almost parallel to each other, whereas the polarization in the slices close to the wall have progressive relative tilts. As a result, the activation barrier Ea for the flipping motion of group of ions in the regions close to the domain-wall would most likely be lower as compared to that for the region deep inside the domain. As the flip probability at any temperature T would vary as exp(-Ea/kT). It is implied that at lower temperatures, when the flips in the interior body of domain would have almost ceased the groups near or inside the wall may still be executing some flipping motions. This in turn, implies that nuclei near the walls would be still experiencing relaxation whereas those deep inside the domain would not be relaxing. Further, it has been recently found by Blinc and coworkers [52] that $^{23}$Na spin-lattice relaxation rate in micro confined $NaNO_2$ in the ferroelectric is similar to that in the bulk. Therefore, we liberally assume that the nature of quadrupolar relaxation for nuclei at the domain-wall is similar to that for those deep inside the domain for order-disorder ferroelectrics in general. Therefore we further assume that the populations at the domain-wall follow the time dependence [53, 54]

$$N_{\pm 1}(0, t) = \alpha e^{-2W_1 t}$$

$$N_0(0, t) = \alpha \left[e^{-2W_1 t} + e^{-2W_2 t}\right] \quad (7)$$

where $W_1$ and $W_2$ are the quadrupolar relaxation probabilities corresponding to the transition m = ±3/2 ↔ ±1/2 and m = ± 3/2 ↔ +1/2 respectively. It should be noted in deciding the location of the origin, i.e. x = 0, that the NMR of nuclei lying inside the wall would not be usually observable due to structural disorders. So, x = 0 would correspond to the region near the wall. At present we are assuming that the wall thickness is negligible.

Using the above boundary conditions, eq. (5) were solved to yield

$$Z_{+1}(x,s) = - a_3\left(1-\frac{\varphi_1^2}{18}\right)e^{-m_1x} - a_4\left(1-\frac{\varphi_2^2}{18}\right)e^{-m_2x} - \frac{A_2}{2}e^{-x\sqrt{\mu s}} + \frac{-k_2+k_3}{2s}$$

$$Z_0(x,s) = a_3e^{-m_1x} + a_4e^{-m_2x} - \frac{k_1-4k_2+k_3}{4s}$$

$$Z_{-1}(x,s) = - a_3\left(1-\frac{\varphi_1^2}{18}\right)e^{-m_1x} - a_4\left(1-\frac{\varphi_2^2}{18}\right)e^{-m_2x} + \frac{A_2}{2}e^{-x\sqrt{\mu s}} + \frac{k_1-k_2}{2s}$$

(8)

where

$$\varphi_1^2 = 25+\sqrt{337}\,, \qquad \varphi_2^2 = 25-\sqrt{337}$$

$$m_1 = \varphi_1\sqrt{\frac{s}{72|b|}}\,, \qquad m_2 = \varphi_2\sqrt{\frac{s}{72|b|}}$$

$$\mu = \frac{2}{9|b|}$$

$$A_2 = -\frac{k_1-k_3}{2s}$$

$$a_3 = -\frac{18}{\varphi_2^2-\varphi_1^2}\left\{\begin{array}{l} -\dfrac{k_1-2k_2+k_3}{4s} - \left(\dfrac{\varphi_1^2}{18}-1\right)\left(\dfrac{k_1-4k_2+k_3}{4s}\right) \\ + \dfrac{\varphi_1^2}{18}\dfrac{\alpha}{s+2W_1} + \left(\dfrac{\varphi_1^2}{18}-1\right)\dfrac{\alpha}{s+2W_2} \end{array}\right\} + \frac{k_1-4k_2+k_3}{4s} - \frac{\alpha}{s+2W_1} - \frac{\alpha}{s+2W_2}$$

$$a_4 = \frac{18}{\varphi_2^2 - \varphi_1^2} \left\{ \begin{array}{l} -\dfrac{k_1 - 2k_2 + k_3}{4s} - \left(\dfrac{\varphi_1^2}{18} - 1\right)\left(\dfrac{k_1 - 4k_2 + k_3}{4s}\right) \\ + \dfrac{\varphi_1^2}{18}\dfrac{\alpha}{s + 2W_1} + \left(\dfrac{\varphi_1^2}{18} - 1\right)\dfrac{\alpha}{s + 2W_2} \end{array} \right\}$$

By taking the inverse Laplace Transform [51] of eq. (8) the values of $N_{\pm 1}$(x, t) and $N_0$(x, t) can be written as

$$N_{+1}(x, t) = C_{11}\,\phi_1(x, t) + C_{12}\,\phi_2(x, t) + C_{13}\,\phi_3(x, t) + C_{14}\,\phi_4(x, t) + C_{15}\,\phi_5(x, t) + C_{16}\,\phi_6(x, t) + C_{17}\,\phi_7(x, t) + C_{18}$$

$$N_{-1}(x, t) = C_{21}\,\phi_1(x, t) + C_{22}\,\phi_2(x, t) + C_{23}\,\phi_3(x, t) + C_{24}\,\phi_4(x, t) + C_{25}\,\phi_5(x, t) + C_{26}\,\phi_6(x, t) + C_{27}\,\phi_7(x, t) + C_{28}$$

$$N_0(x, t) = C_{31}\phi_1(x, t) + C_{32}\,\phi_2(x, t) + C_{33}\,\phi_3(x, t) + C_{34}\,\phi_4(x, t) + C_{35}\,\phi_5(x, t) + C_{36}\,\phi_6(x, t) + C_{38} \qquad (9)$$

where

$$C_{11} = C_{21} = -\left(1 - \frac{\varphi_1^2}{18}\right) C_{31}$$

$$= -\left(1 - \frac{\varphi_1^2}{18}\right) \left[ \begin{array}{l} \dfrac{\frac{18}{4}\left(k_1 - 2k_2 + k_3\right)}{\varphi_2^2 - \varphi_1^2} + \dfrac{18}{\varphi_2^2 - \varphi_1^2} \times \left(\dfrac{\varphi_1^2}{18} - 1\right) \\ \dfrac{\left(k_1 - 4k_2 + k_3\right)}{4} + \dfrac{\left(k_1 - 4k_2 + k_3\right)}{4} \end{array} \right]$$

$$C_{12} = C_{22} = -\left(1 - \frac{\varphi_1^2}{18}\right) C_{32} = \left(1 - \frac{\varphi_1^2}{18}\right)\left(\frac{\varphi_2^2}{\varphi_2^2 - \varphi_1^2}\right)$$

$$C_{13} = C_{23} = -\left(1 - \frac{\varphi_1^2}{18}\right) C_{33} = \left(1 - \frac{\varphi_1^2}{18}\right)\left[\left(\frac{\varphi_1^2}{18} - 1\right)\frac{18}{\varphi_2^2 - \varphi_1^2} + 1\right]$$

$$C_{14} = C_{24} = -\left(1-\frac{\varphi_2^2}{18}\right) C_{34}$$

$$= -\left(1-\frac{\varphi_2^2}{18}\right)\left[\begin{array}{c}\dfrac{-\frac{18}{4}\left(k_1 - 2k_2 + k_3\right)}{\varphi_2^2 - \varphi_1^2} - \dfrac{18}{\varphi_2^2 - \varphi_1^2}\times \\ \left(\dfrac{\varphi_1^2}{18} - 1\right)\left(\dfrac{\left(k_1 - 4k_2 + k_3\right)}{4}\right)\end{array}\right]$$

$$C_{15} = C_{25} = -\left(1-\frac{\varphi_2^2}{18}\right) C_{35} = \left(1-\frac{\varphi_2^2}{18}\right)\left(\frac{\varphi_1^2}{\varphi_2^2 - \varphi_1^2}\right)$$

$$C_{16} = C_{26} = -\left(1-\frac{\varphi_2^2}{18}\right) C_{36} = \left(1-\frac{\varphi_2^2}{18}\right)\left[\left(\frac{\varphi_1^2}{18} - 1\right)\frac{18}{\varphi_2^2 - \varphi_2^2}\right]$$

$$C_{17} = -\frac{k_1 - k_3}{4}, \quad C_{27} = \frac{k_1 - k_3}{4}, \quad C_{18} = \frac{k_1 - k_2}{4}, \quad C_{28} = \frac{k_3 - k_2}{4},$$

$$C_{38} = -\frac{k_1 - 4k_2 + k_3}{4}$$

and

$$\phi_1(x,t) = erfc\left(\frac{x\varphi_1}{\sqrt{72|b|}}\frac{1}{2\sqrt{t}}\right)$$

$$\phi_2(x,t) = \frac{\alpha}{2}\, e^{-2W_1 t}\left\{\begin{array}{l} \mathrm{e}^{-\sqrt{-2W_1}\frac{x\varphi_1}{\sqrt{72|b|}}}\; erfc\left(-\sqrt{-2W_1 t} + \dfrac{x\varphi_1}{\sqrt{72|b|}}\dfrac{1}{2\sqrt{t}}\right) \\ + \;\mathrm{e}^{\sqrt{-2W_1}\frac{x\varphi_1}{\sqrt{72|b|}}}\; erfc\left(\sqrt{-2W_1 t} + \dfrac{x\varphi_1}{\sqrt{72|b|}}\dfrac{1}{2\sqrt{t}}\right)\end{array}\right\}$$

$$\phi_3(x,t) = \frac{\alpha}{2}\, e^{-2W_2 t}\left\{\begin{array}{l} \mathrm{e}^{-\sqrt{-2W_2}\frac{x\varphi_1}{\sqrt{72|b|}}}\; erfc\left(-\sqrt{-2W_2 t} + \dfrac{x\varphi_1}{\sqrt{72|b|}}\dfrac{1}{2\sqrt{t}}\right) \\ + \;\mathrm{e}^{\sqrt{-2W_2}\frac{x\varphi_1}{\sqrt{72|b|}}}\; erfc\left(\sqrt{-2W_2 t} + \dfrac{x\varphi_1}{\sqrt{72|b|}}\dfrac{1}{2\sqrt{t}}\right)\end{array}\right\}$$

$$\phi_4(x,t) = erfc\left(\frac{x\varphi_2}{\sqrt{72|b|}}\;\frac{1}{2\sqrt{t}}\right)$$

$$\phi_5(x,t) = \frac{\alpha}{2} e^{-2W_1 t} \left\{ \begin{array}{l} e^{-\sqrt{-2W_1}\frac{x\varphi_2}{\sqrt{72|b|}}} erfc\left( -\sqrt{-2W_1 t} + \frac{x\varphi_2}{\sqrt{72|b|}}\frac{1}{2\sqrt{t}} \right) \\ + \; e^{\sqrt{-2W_1}\frac{x\varphi_2}{\sqrt{72|b|}}} erfc\left( \sqrt{-2W_1 t} + \frac{x\varphi_2}{\sqrt{72|b|}}\frac{1}{2\sqrt{t}} \right) \end{array} \right\}$$

$$\phi_6(x,t) = \frac{\alpha}{2} e^{-2W_2 t} \left\{ \begin{array}{l} e^{-\sqrt{-2W_2}\frac{x\varphi_2}{\sqrt{72|b|}}} erfc\left( -\sqrt{-2W_2 t} + \frac{x\varphi_2}{\sqrt{72|b|}}\frac{1}{2\sqrt{t}} \right) \\ + \, e^{\sqrt{-2W_2}\frac{x\varphi_2}{\sqrt{72|b|}}} erfc\left( \sqrt{-2W_2 t} + \frac{x\varphi_2}{\sqrt{72|b|}}\frac{1}{2\sqrt{t}} \right) \end{array} \right\}$$

$$\phi_7(x,t) = erfc\left( x\sqrt{\frac{2}{9|b|}}\frac{1}{2\sqrt{t}} \right)$$

The functions $\phi_2(x, t)$, $\phi_3(x, t)$, $\phi_5(x, t)$ and $\phi_6(x, t)$ can be simplified using the numerical expansion of complex error function (See Appendix- I)

The average values of the population differences for the entire domain would then be given by

$$N_i(t) = \frac{1}{L}\int_0^L N_i(x,t)\, dx \tag{10}$$

where $i$ =1, 0 or –1 and L is the thickness of the domain. In our treatment domain-wall thickness has been ignored. For clarity the symbols and useful expressions have been kept same as taken in the Ref. [47].

## 4. Result and Discussion

The time dependence of the population differences $N_{+1}(t)$, $N_{-1}(t)$ and $N_0(t)$ were evaluated numerically using eq.(9) and (10) for different values of the ratios $W_1/W_{00}$ taking $W_2/W_1 = 1$ and various values of L/a. A typical plot is shown in figure 3 for L/a = 5 (i.e. 1.78 nm for $NaNO_2$, wherever needed we have taken relevant data for ferroelectric $NaNO_2$ just to see the behaviour). The curves marked 1, 2, 3, 4, 5 correspond to the values $W_1/W_{00}$ = 1, 0.1, 0.001, 0.001 and 0.0001 respectively whereas the curve 6 represents the behaviour given by eq. (7) i.e. the case where the nuclei at all the sites x throughout the domain follow the relation given by eq. (7) and spin diffusion is absent. It is seen that for all the cases, $N_{\pm 1}(t)$ are in general non exponential. It may be mentioned that it was shown by Kotecha and Pandey [47] that values of $W_2/W_1$ in the range of 0.5 to 2.0 which are typical for ferroelectrics such as $NaNO_2$ do not affect the population difference much. Therefore results were obtained for $W_2/W_1 = 1$ only. From figure 3 it is seen that $N_{\pm 1}(t)$ for all these values are non exponential. If we treat the quantity $(W_{00}a^2)^{1/2}$ as the diffusion coefficient D, then the function $(W_{00}a^2t)^{1/2}$ becomes the diffusion length and provides an estimate of the distance up to which the magnetization would have diffused from the domain-walls (x = 0) into the domain in time t. The quantity $(L/\sqrt{W_{00}a^2t})$ then gives an estimate of the portion of the domain of length L getting affected in time t due to relaxation occurring in the domain wall. Values of the population differences $N_{\pm 1}(t)$ after a $\pi/2$ pulse (i.e. $\alpha = 0.5$) were calculated for various values of L/a and were plotted. Figure 4 shows a typical plot of $\log_e N_{\pm 1}(t)$ vs $\log_e [\sqrt{W_{00}t} / (L/a)]$ for L/a= 5 i.e. L = 1.78 nm and L/a=30 i.e. L = 10.68nm for $NaNo_2$ (a = 3.56 $A^0$).

This plot shows the changes in population differences as predicted by eq. (9) and (10) i. e. affected due to domain walls. However, these changes are not easy to visualize from eq. (9) and (10). For an easy visualization of the relaxation proceeding in the domain, an attempt was made to represent the behaviour shown in figure 4 by some empirical relation. From the figure 4 we find that for the values of $\log_e [\sqrt{W_{00}t} / (L/a)] < -3$ i.e. $\sqrt{W_{00}a^2t} < L \exp(-3)$ i.e. $\sqrt{W_{00}a^2t} < 0.05L$ the graph is flat indicating that the value of $\log_e N_{\pm 1}(t)$ is constant and is equal to - 0.6931 giving $N_{\pm 1}(t) = 0.5$ which is nothing but the initial population difference created after the pulse. It means that for times shorter than $0.05\ L / \sqrt{W_{00}a^2}$ the effect of domain-wall on the domain would not be visible and the value of $N_{\pm 1}(t)$ can be taken as $N_{\pm 1}(0)$. Similarly, from figure 4 we see that for values of $\log_e[\sqrt{W_{00}t} / (L/a)]$ greater than 2.5 the plot is a straight line with a negative slope indicating a power law dependence. This linear relation can be written as

$$\log_e N_{\pm 1}(t) = - p \log_e[\sqrt{W_{00}t} / (L/a)] + c$$

where –p is the slope and c is the intercept for the line. From this we find that $N_{\pm1}(t)$ follows the relation

$$N_{\pm1}(t) = C_1\ (L/a)^p(1/W_{00})^{p/2}\ t^{-p/2}.$$

The values of $C_1$ and p would depend upon the ratios $W_2/W_1$ and $W_1/W_{00}$. For example, we get p=1.03 and $C_1$=exp(-2) for $W_2/W_1 = 1$ and $W_1/W_0 = 1$. The behaviour near the value $\log_e[\surd W_{00}t\ /\ (L/a)] = 0$ is given by curved portion of the graph.. It was found that the overall behaviour of $N_{\pm1}(t)$ shown in figure 4 and governed by eq. (10) may be crudely expressed by the empirical relation

$$N_{\pm1}(t) = \frac{N_{\pm1}(0)}{\left[1 + d\left(\frac{\sqrt{W_{00}t}}{L/a}\right)^p\right]} \tag{11}$$

where p and d depend upon the ratio $W_1/W_{00}$. Their values were found not to vary significantly with the ratio $W_2/W_1$ and lie in the range 1 to 3 and 0.4 to 3 respectively. The calculated value of p and d for different values of L/a are given in Table 1. The eq. (11) represents the time dependence of the magnetization corresponding to the transitions $\pm3/2\leftrightarrow\pm1/2$ as the population would be proportional to $N_{\pm1}(t)$. If we take the relaxation time as the duration in which $N_{\pm1}(t)$ has decayed to $e^{-1}$ of the initial value $N_{\pm1}(0)$, we find that the relaxation time say $T_{1dw}$ due to domain wall effect can be written as

$$T_{1dw} = \frac{1}{W_{00}} \frac{\left(\frac{L}{a}\right)^2}{\left(\frac{d}{e-1}\right)^{\frac{2}{p}}} \tag{12}$$

Similar treatment may be carried out for the behaviour of $N_0$ (t).

The relaxation rate ($1/T_{1dw}$) due to domain wall given by eq. (12) is plotted as a function of (L/a) in figure 5. It is clear from figure 5 that $1/T_{1dw}$ is large for very small values of L/a, decreases very fast and reaches almost a constant value for large values of (L/a). Here L is the domain width and 'a' is the inter-nuclear spacing. From the figure 5 it is indicated that, if the spin lattice relaxation time $T_1$ for a ferroelectric of usual domain size is known, then the domain width of an unknown sample of the same ferroelectric can be estimated by measuring the value of $T_1$ for that sample and using figure 5 or eq. (12). These results are quite general and can be applied to any ferroelectric system having $180^{o}$ domains. It is worth mentioning here that, as shown in figure 5, $1/T_{1dw}$ shows a noticeable variation only for smaller values of L/a (δ20) suggesting that this method would be useful only for the samples with domain width δ10 nm (for a = $3.56A^{0}$). For larger domain widths this method would not work. Also it should be mentioned here that spin lattice relaxation for I=3/2 system in a ferroelectric would get contributions from various mechanisms. However as mentioned earlier, in most of the ferroelectrics the relaxation is usually quadrupolar and proceeds through the interaction between nuclear electric quadrupole moment and the fluctuating electric field gradient which generally arises due to flipping/tumbling motions of groups (for example $NO_2$ group in $NaNO_2$ ). As these flipping/tumbling motions almost cease at low temperatures the relaxation rate becomes very small at low temperatures in a quadrupolar system. Thus the method that we propose here involves two steps: first the spin lattice relaxation time $T_1$ of I=3/2 nuclei in the ferroelectric material should be measured at low temperatures, second the value of $T_1$ is measured again at same temperature for the same nuclei but in the sample with unknown nano sized domains. Then the width of the nano domains is obtained from the figure 5 just by comparison and eq. (12). The values of d, p and $W_{00}$ appearing in eq. (12) would be constant for a given material and would in general be different for different materials. Therefore we would have specific graph such as figure 5 for different materials. These results are general and are expected to prompt experimentalists to verify them.

## 4. Conclusions

The spin-lattice relaxation of I=3/2 quadrupolar spin system due to domain-walls in order-disorder ferroelectric was theoretically studied by representing the $180^{0}$ domain by a chain of equidistant I=3/2 spins. The electric polarization undergoes spiral orientation as one moves one domain to the adjacent one. This implies that if we treat the domain to be made up of thin slices, then the polarization in the slices deep inside the domain would be almost parallel to each other whereas the polarization in slices near the domain-wall would undergo larger relative tilts. Therefore owing to the lower activation barrier near the walls the probability of activated reorientational flipping motions of group of ions near the wall would be more. As a result at a given temperature the nuclei near the walls would be experiencing quadrupolar relaxation whereas those inside the domain would not do so. This would lead to spin diffusion from the domain-walls. Rate equations are formed for the population difference for $I=3/2$ quadrupolar nuclei and are solved analytically using Laplace Transform. Expression for spin-lattice relaxation rate due to domain-walls is derived in terms of domain width. A general method is proposed for estimation of domain width of a nano-ferroelectric by measuring the spin-lattice relaxation time $T_1$ at low temperatures and using the $T_1$ data for the same ferroelectric having domains of usual micron sizes.

## Appendix- I

The Laplace transform of any function $f$(t) is defined as [45,46].

$$\mathrm{F(s)} = \mathcal{L}[f(\mathrm{t})] = \int_0^{\infty} \mathrm{e}^{-\mathrm{st}} f(\mathrm{t})\, \mathrm{dt}$$

where s is the Laplace variable. $f$(t) is called the original function and F(s) is called the image function.

The error function is a special function and defined as [46]

$$erf(z) = \frac{2}{\pi}\int_0^z e^{-t^2}\, dt$$

$$erfc(z) = \frac{2}{\pi}\int_z^{\infty} e^{-t^2}\, dt = 1 - erf(z)$$

Numerical Approximation for complex error function is given by

$$erf(x+iy) = erf(x) + \frac{e^{-x^2}}{2\pi x}\left[(1-\cos 2xy) + i\sin 2xy\right] + \frac{2}{\pi}e^{-x^2}\sum_{n=1}^{\infty}\frac{e^{\frac{-1}{4}n^2}}{n^2+4x^2}\left[f_n(x,y) + ig_n(x,y)\right] + \in(x,y)$$

*where*

$$f_n(x,y) = 2x - 2x\cosh ny.\cos 2xy + n\sinh ny.\sin 2xy$$

$$g_n(x,y) = 2x\cosh ny.\sin 2xy + n\sinh ny.\cos 2xy$$

$$\left|\in(x,y)\right| \approx 10^{-16}\left|erf(x+iy)\right|$$

**Figure 1**

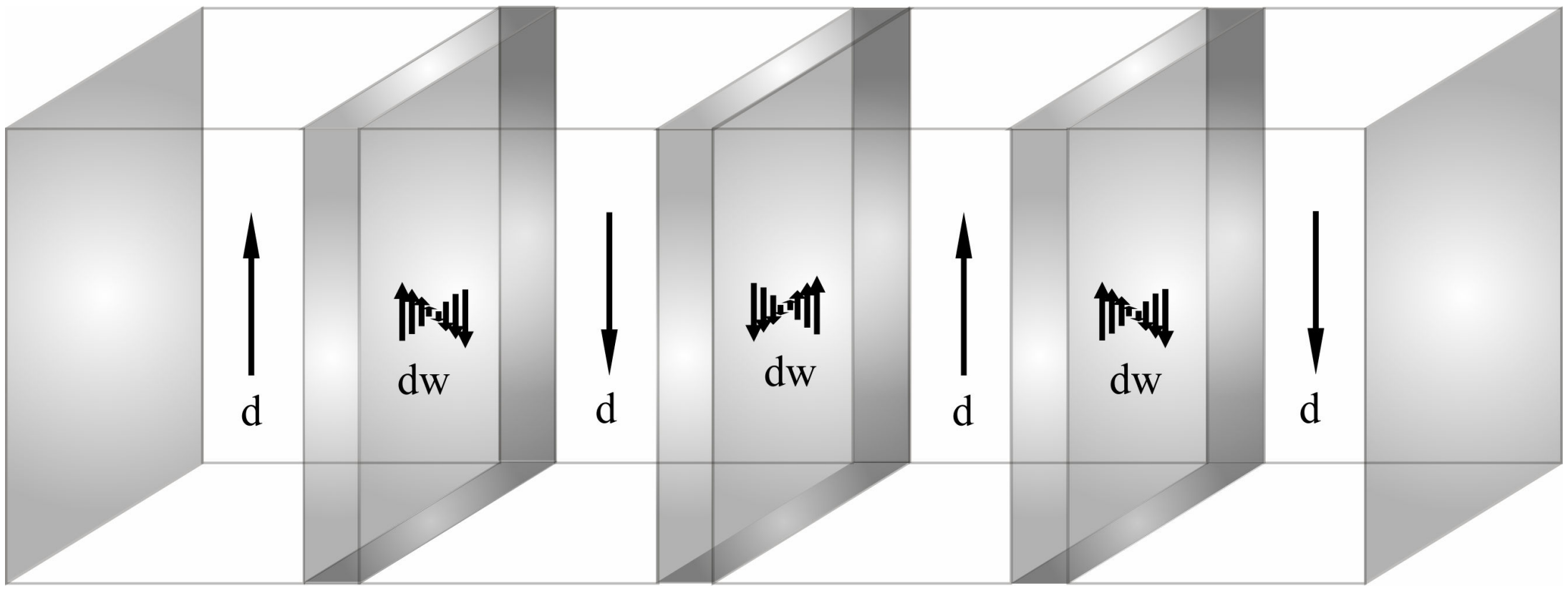

Figure 1: Schematic diagram of 180° domain domain-wall structure.

**Figure 2**

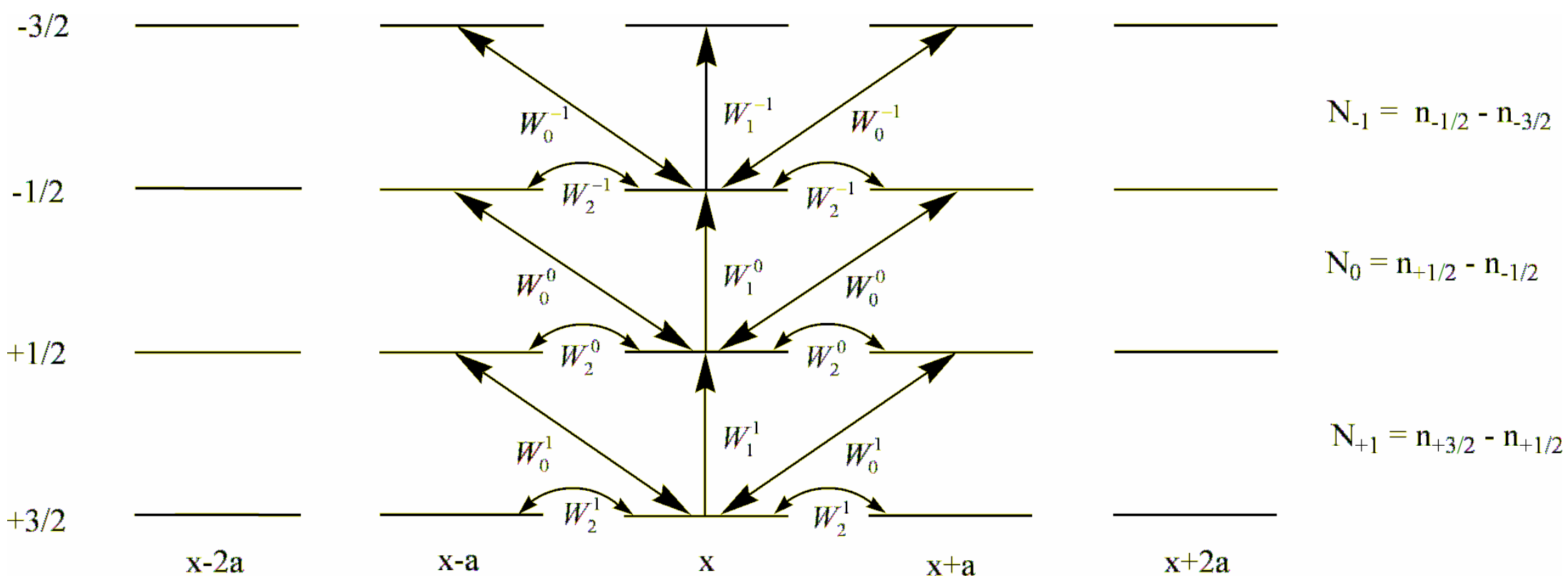

Figure 2: Schematic energy level diagrams of spins (I= $^3/_2$) in a one - dimensional chain. $W_2^{1}$, $W_2^{0}$, $W_2^{-1}$ are the transition probabilities of $^3/_2 \leftrightarrow ½$ ½ $\leftrightarrow$½, ½ $\leftrightarrow ^{-3}/_2$ levels respectively for the case where one spin undergoing an upward transition while the other spin undergoes downward transition (usually called flip-flop term). $W_2^{1}$, $W_2^{0}$, $W_2^{-1}$ represent simultaneous upward (or down ward) flip of the pair of spins. $W_2^{1}$, $W_2^{0}$, $W_2^{-1}$ represent the single spin transition probability for the spin pairs and counted twice for each pair of spins.

**Figure 3**

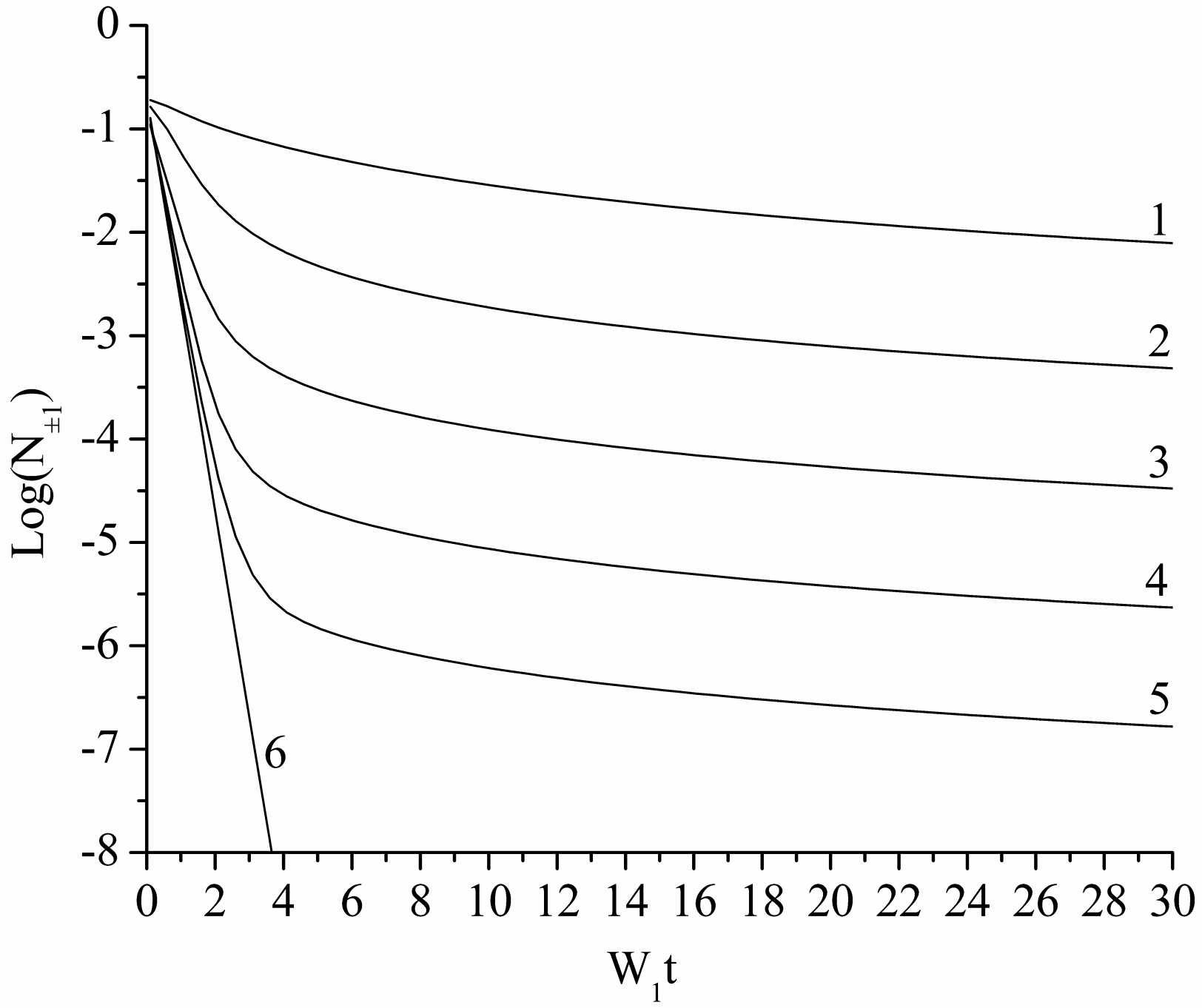

Figure 3: Variation of $\log_e N_{\pm 1}$ (t) as a function of $W_1 t$ for different ratios of W1/$W_{oo}$ using $W_2$ / $W_1 = 1$ and L/a = 5 with curve number 1, 2, 3, 4, 5 representing $W_1$ / $W_{oo}$ = 1, 0.1, 0.01, 0.001, 0.0001 respectively. The curve 6 represents the behaviour given by eq. (7).

**Figure 4**

**(a)**

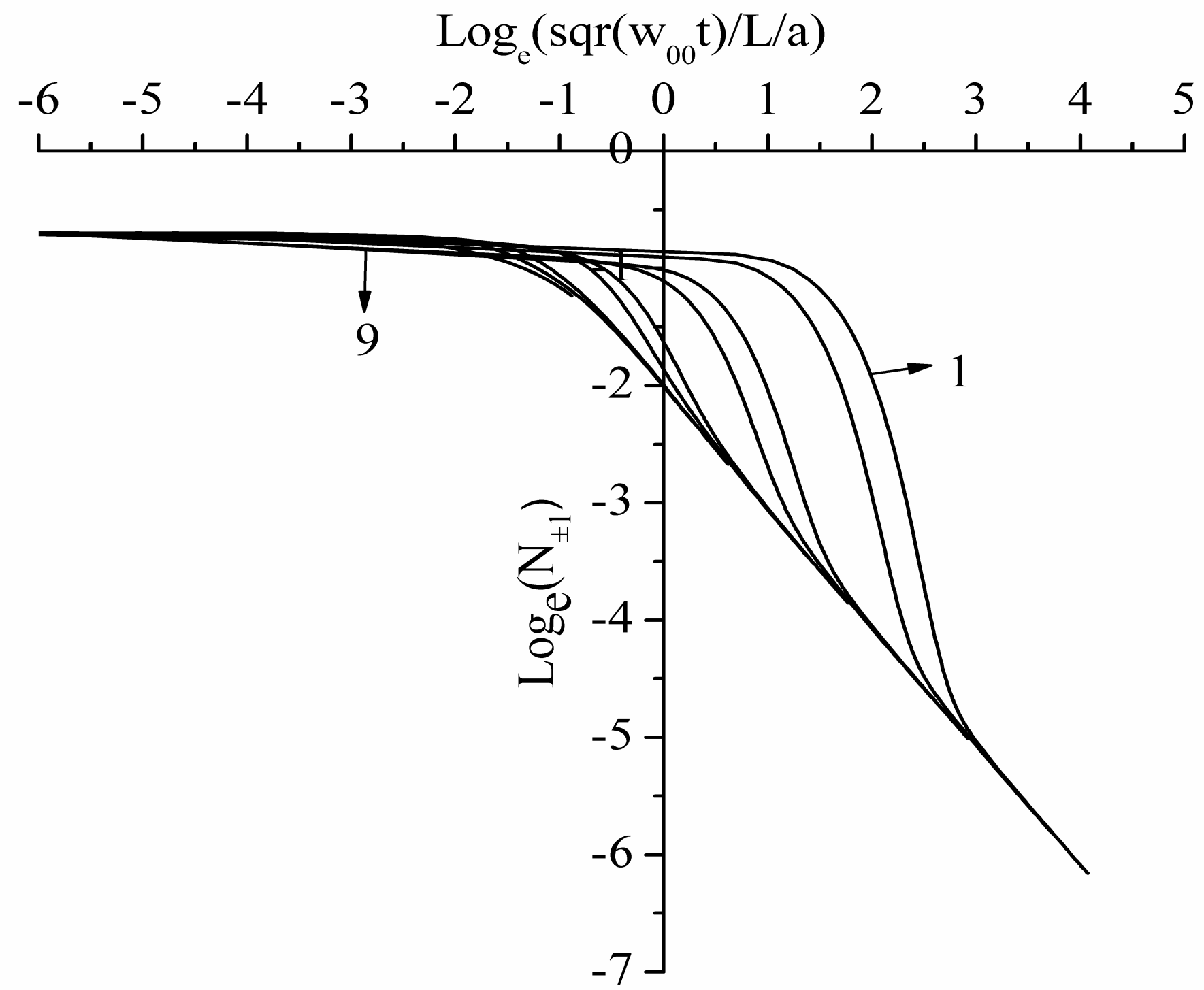

**(b)**

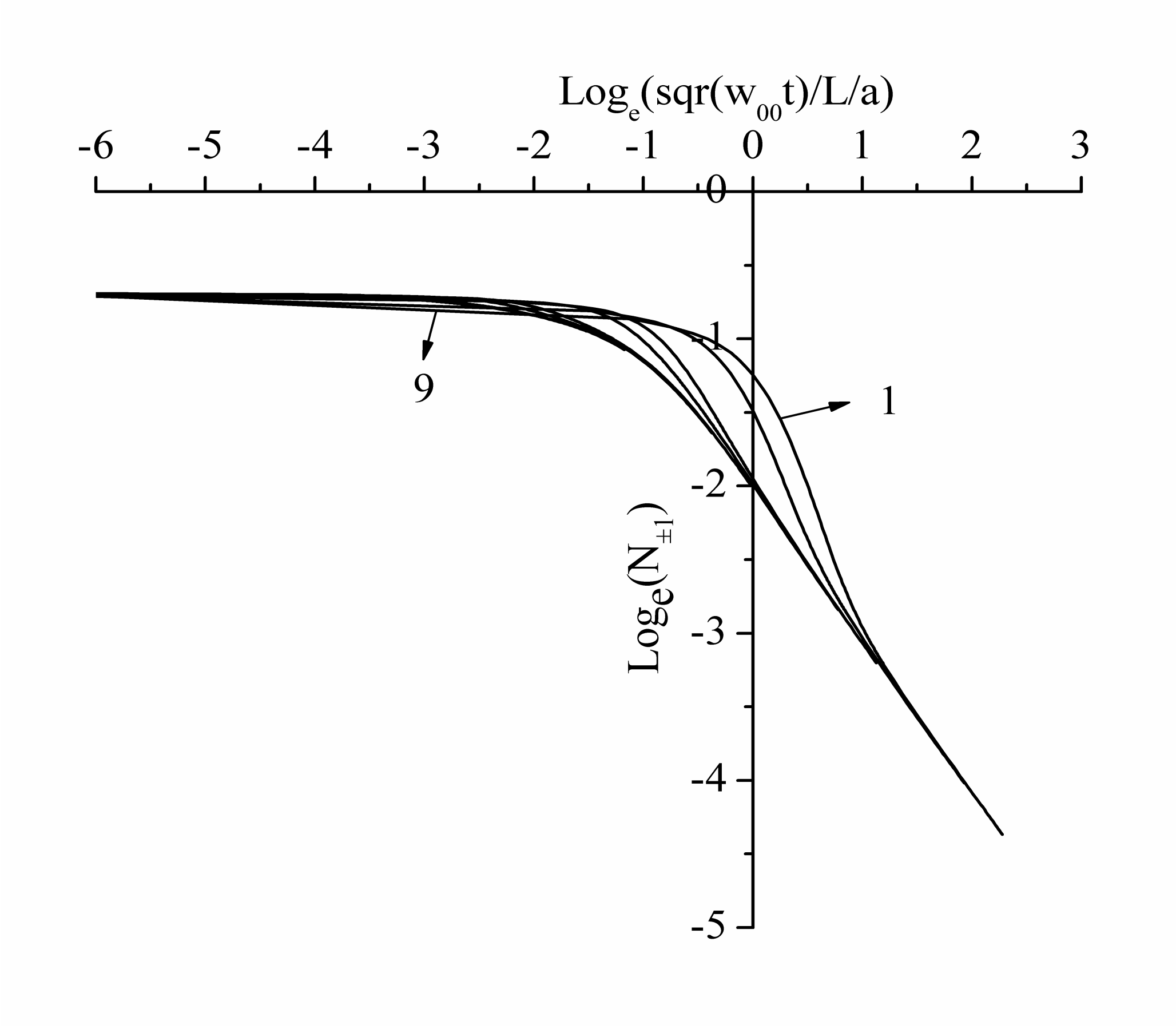

Figure 4: Variation of $\log_e N_{\pm 1}(t)$ as a function of $\mathrm{Log}_e\sqrt{(W_{oo}t)}/L/a$ for different $W_1/W_{oo}$ ratio using **(a)** L/a= 5 (L = 1.78 nm) and **(b)** L/a=30 (L = 10.68 nm). Curve no. 1 through 9 represent $W_1/W_{oo}$ = 0.0005, 0.001, 0.005, 0.01, 0.05, 0.1, 0.5, 1 and 10 respectively.

**Figure 5**

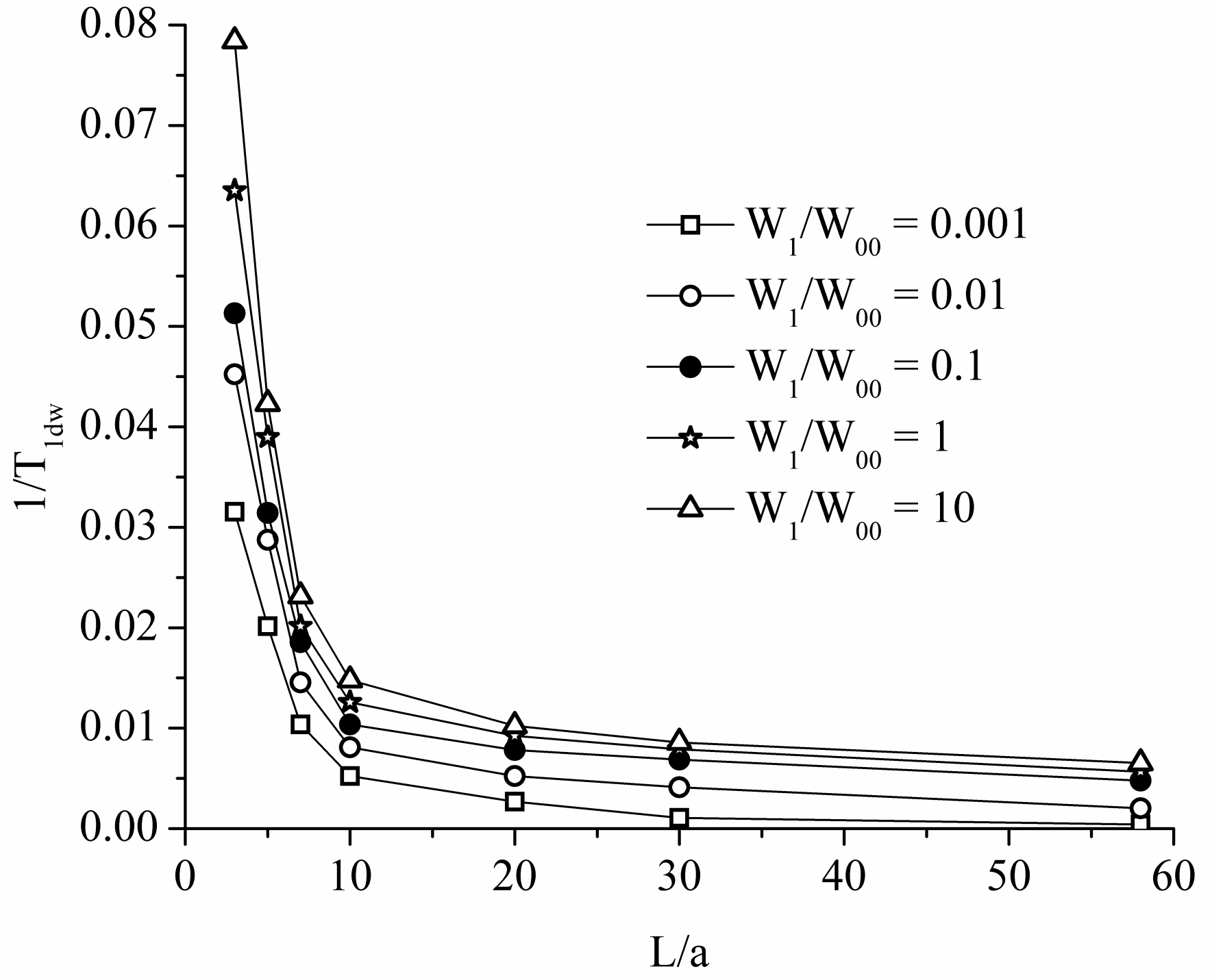

Figure 5: The relaxation probability ($1/T_{1dw}$) as a function of ratio (L/a) for $W_1/W_{oo}$ = 0.001, 0.01, 0.1, 1 and 10. The value of d and p used are given in Table 1.

**Table 1:** The value of p and d for different ratio of L/a as obtained from equation 11.

| L/a (L) | w1/w00=0.001 | | w1/w00=0.01 | | w1/w00=0.1 | | w1/w00=1 | | w1/w00=10 | |
|---|---|---|---|---|---|---|---|---|---|---|
| | **p** | **d** | **p** | **d** | **p** | **d** | **p** | **d** | **p** | **d** |
| 3 (1.068nm) | 2.93 | 0.43 | 2.53 | 0.53 | 2.03 | 0.87 | 1.68 | 1.13 | 1.43 | 2.17 |
| 5 (1.78nm) | 2.71 | 0.58 | 2.27 | 0.69 | 1.94 | 0.93 | 1.52 | 1.74 | 1.27 | 2.48 |
| 7 (2.49nm) | 2.52 | 0.74 | 2.01 | 0.86 | 1.72 | 1.47 | 1.31 | 2.36 | 1.19 | 2.82 |
| 10 (3.56nm) | 2.31 | 0.89 | 1.92 | 1.12 | 1.54 | 1.96 | 1.23 | 3.68 | 1.13 | 3.71 |
| 20 (7.12nm) | 2.02 | 1.04 | 1.805 | 1.43 | 1.37 | 2.07 | 1.16 | 3.74 | 1.1 | 3.805 |
| 30 (10.68nm) | 1.97 | 1.13 | 1.72 | 1.68 | 1.24 | 2.46 | 1.1 | 3.86 | 1.09 | 3.91 |
| 58 (20.64nm) | 1.81 | 1.27 | 1.64 | 1.88 | 1.19 | 2.97 | 1.04 | 3.91 | 1.01 | 3.96 |